\documentclass[structabstract]{aa}  

\pdfoutput=0
\usepackage{graphicx}
\usepackage{sidecap}
\usepackage{txfonts}
\usepackage{natbib}

\newcommand{\mk}{} 

\begin{document}

\subtitle{The low surface brightness host of SN~2009Z}

\title{Supernovae without host galaxies?}

\author{P.-C. Zinn\inst{1,2}
  \and
  M. Stritzinger\inst{3,4}
  \and
  J. Braithwaite\inst{5}
  \and
  A. Gallazzi\inst{4}
  \and
  P. Grunden\inst{1}
  \and
  D. J. Bomans\inst{1}
  \and
  N. I. Morrell\inst{6}
  \and
  U. Bach\inst{7}
}

\institute{Astronomical Institute, Ruhr-University Bochum, Universit\"atsstra\ss{}e 150, 44801 Bochum, Germany\\
  \email{zinn@astro.rub.de}
  \and
  CSIRO Astronomy \& Space Science, PO Box 76, Epping, NSW, 1710, Australia
  \and
  The Oskar Klein Centre, Department of Astronomy, Stockholm University, AlbaNova, 10691 Stockholm, Sweden
  \and
  Dark Cosmology Centre, Niels Bohr Institute, University of Copenhagen, Juliane Maries Vej 30, 2100 Copenhagen \O , Denmark
   \and
  Argelander Institut f\"ur Astronomie, Universit\"at Bonn, Auf dem H\"ugel 71, 53121 Bonn, Germany
  \and
  Las Campanas Observatory, Carnegie Observatories, Casilla 601, La Serena, Chile
  \and
  Max-Planck-Institute for Radio Astronomy, Auf dem H\"ugel 69, 53121 Bonn, Germany
  }

\date{Received 04/01/2011; Accepted 07/11/2011}

\abstract
{A remarkable fraction of supernovae (SNe) have no obvious host galaxy. Two possible explanations are that (i) the host galaxy is simply not detected within the
  sensitivity of the available data or that (ii) the progenitor is 
  a hypervelocity star that has escaped its parent galaxy.}
{{\mk We use the Type IIb SN~2009Z as a prototype of case (i), an example of how a 
very faint (here Low Surface Brightness; LSB) galaxy can be discovered via the observation of a seemingly 
host-less SN. By identifying and studying LSB 
galaxies
that host SNe related to the death of massive stars, we can place
constraints on the stellar population and environment of LSB galaxies,
which at present are poorly understood.}}
{\mk{We use archival ultraviolet (UV) and optical imaging,  as well as an H\,I spectrum taken
  with the 100\,m Effelsberg Radio Telescope to measure various parameters of the host galaxy, in particular its redshift, stellar and H\,I mass, and metallicity.}}
{From the Effelsberg spectrum, a redshift $z$\,=\,0.02513$\pm$0.00001 and an H\,I mass of 2.96$\pm$0.12 10$^9\,M_{\odot}$ are computed. This redshift is consistent with that obtained from optical emission lines of SN~2009Z.
Furthermore, a gas mass fraction of $f_g\,=\,0.87\pm0.04$ is obtained, one of the highest fractions ever measured. The host galaxy shows signs of recently enhanced star formation activity with a far-UV derived extinction-corrected Star Formation Rate (SFR) of 0.44$\pm0.34\,M_{\odot}$\,yr$^{-1}$.
Based on the $B$-band luminosity we estimate an extinction-corrected metallicity following the calibration by Pilyugin (2001)
of $12+\log\left(\frac{\rm{O}}{\rm{H}}\right)\,=\,8.24\pm 0.70$.}
{\mk{The presence of a Type IIb SN in an LSB galaxy suggests, contrary to popular belief, that
massive stars can be formed in this type of galaxies.   
Furthermore, our results imply that LSB galaxies undergo phases of small, local burst activity intermittent with longer phases of inactivity, rather than a continuous but very low SFR. 
Discovering faint (LSB) galaxies via bright supernova events happening in them offers an excellent opportunity to improve our understanding of the nature of LSB galaxies.}}

\keywords{supernovae: individual: SN~2009Z --
  galaxies: evolution --
  galaxies: stellar content --
  methods: observational
}

\maketitle

\section{Introduction}
\label{Intro}
The Sternberg Astronomical Institute (SAI) supernova (SN) catalog \citep{sai}
lists over 5000 objects of which a surprising fraction have no obvious host galaxy. 
Two possible scenarios discussed in the
literature \citep[e.g.][]{hayward05} are:
\begin{enumerate}
\item The host galaxies are simply not detected, given
  the sensitivity of  the available data;
\item The progenitors of these SNe are
  hypervelocity stars \citep[$v\,\geq\,100$\,km\,s$^{-1}$, see][]{martin06} that have escaped the gravitational potential of their parent galaxy.
\end{enumerate}

In this paper we examine the case of the Type IIb SN~2009Z, which is an example of possibility 1. 
The nearest possible host appeared initially to be the face-on spiral galaxy UGC~8939, whose core
lies nearly 3\,arcmin away, i.e. $\sim90\,$kpc on the plane of the sky, from the location of the SN.
However, under close inspection of deep archival images, an irregular dwarf galaxy, 2dFGRS\,N271Z016 \citep[hereafter N271, also known as J140153.80-012035.5 in the Sloan Digital Sky Survey; SDSS][]{SDSSDR7}, has been identified as the true host. This galaxy is at the edge of SDSS detection limits -- the SDSS designation was assigned to it prior to the supernova, but the detection is at a very low confidence limit.
We classify this galaxy below as a low surface brightness (LSB) galaxy
 \citep[for a concise review of this
class of galaxies see for example][]{impeybothun}. 
 This is interesting, since SNe are rarely found in LSB galaxies.

As easily detectable point sources, SNe are a promising tool for discovering very faint and/or LSB galaxies. In contrast, sensitivity-limited galaxy surveys yield an incomplete sample of galaxies in which LSB galaxies are very likely to be underrepresented. Consequently, the contribution of LSB galaxies to both the total baryon density of the universe and their contribution to the galaxy number density are still uncertain.
For example, \cite{hayward05} argued that LSB galaxies contain only a small fraction of the baryons and are therefore `cosmologically unimportant', 
 whereas \cite{Minchin04} found that LSB galaxies account for $62\pm37\%$ of gas-rich galaxies by number. 


Obviously to use SNe to find faint galaxies, a survey needs to be `non-targeted' rather than looking at likely SN locations. \mk{Currently there are several wide-field, non-targeted SN surveys underway \citep[e.g. Pan-STARRS or the Palomar Transient Factory,][]{PanStarrs,PTF} whose goal, amongst others, is to characterize SN events that occur in all types of galaxies. SN~2009Z in N271 can therefore be put into context with a number of other recently studied core-collapse (CC) SNe, including those that are associated with a long-duration gamma-ray burst (GRB), that have occurred in faint dwarf galaxies. A number of long GRBs have now been associated with broad-lined Type Ic SNe, but it is clear from the statistics that not {\it all} broad-lined Type Ic SNe produce a GRB, either on-axis or off-axis. Long GRBs are found preferentially in small irregular galaxies, and in the more luminous parts of their hosts, in this respect similar to Type Ic SNe, but unlike Type II SNe \citep{Fruchter2006,Kelly2008}. Furthermore, \citet{Modjaz2008} found that hosts of Type Ic SNe associated with a GRB have lower metallicity on average than those without any GRB. In general long GRBs are associated with low metallicity \citep{Stanek2006}, which may be related to the bias towards high redshift. From a theoretical point of view, is it thought that low metallicity somehow helps the progenitor to retain more angular momentum, via suppression of wind, for instance. It is generally accepted that rapid core rotation is essential to produce a GRB, as well as a progenitor significantly above the $\sim8$ $M_\odot$ CCSN threshold \citep[see e.g. ][ and refs.\ therein]{Woosley2011}. In any case, the matter of SNe~Ic with and without GRBs clearly demonstrates the need for more thorough studies of SNe {\it and} their hosts, in particular a larger variety of hosts -- previous SN surveys have targeted mainly bright, giant galaxies.

Although all the work summarized above mainly focusses on Type~Ic events and associated GRBs, other types of CCSNe are useful in shedding light on the stellar population of their host galaxies, since they also require high-mass progenitors. For instance, studies of the environments of regular and stripped CCSNe have been made on their metallicity \citep{Anderson2011,Modjaz2011,Leloudas2011} and the age of the stellar population \citep{Leloudas2011}.  

LSB galaxies, according to prevailing opinion, have
 a comparable total H\,I mass to high surface brightness (HSB) galaxies, but a lower surface density, too low for molecular clouds to form \citep[for the surface density criterion see][]{Kennicutt89}. This leads to lower rates of star formation and metal production. Not surprisingly, LSB galaxies have higher mass-to-light ratios than HSB galaxies \citep{deBlok96}.

In this paper we present our examination of the properties of N271, the LSB host galaxy of SN~2009Z, concentrating on its stellar population, gas mass fraction and the other properties of its interstellar medium (ISM).
Throughout this paper, we adopt a flat $\Lambda$CDM cosmology with $H_0\,=\,70.2$\,km\,s$^{-1}$\,Mpc$^{-1}$ and $\Omega_{\Lambda}$\,=\,0.73 \citep{Komatsu2011}.
 
In the next section, we describe the observations of the supernova and its host, both archival and current. In section \ref{properties} we look at the properties of the host galaxy in detail, before discussing the results in the context of other recent work in section \ref{discussion} and summarising in section \ref{conc}.
 
\section{Observations}

\subsection{Supernova 2009Z}
SN~2009Z was discovered on 2.53 February 2009 UT by the 
the Lick Observatory Supernova Search \citep{filippenko01} with an unfiltered magnitude of 18.1. Soon afterwards \cite{cbet1703} 
classified it as a Type IIb, 
spectroscopically most similar to SN~1993J around maximum.
Detailed optical and near-IR observations were obtained by the 
Carnegie Supernova Project \citep{Hamuy06}. An analysis of 
preliminary light curves reveals a peak $B$-band maximum 
of 17.85$\pm$0.10 on 13.8 February 2009. 
Adopting a distance of 108 Mpc 
 (see sec \ref{HI}), this corresponds to  
an absolute $B$-band magnitude of $M_{B}\,=\,-17.32\pm 0.15$.
\begin{figure*}
  \centering
  \includegraphics[width=0.49\textwidth]{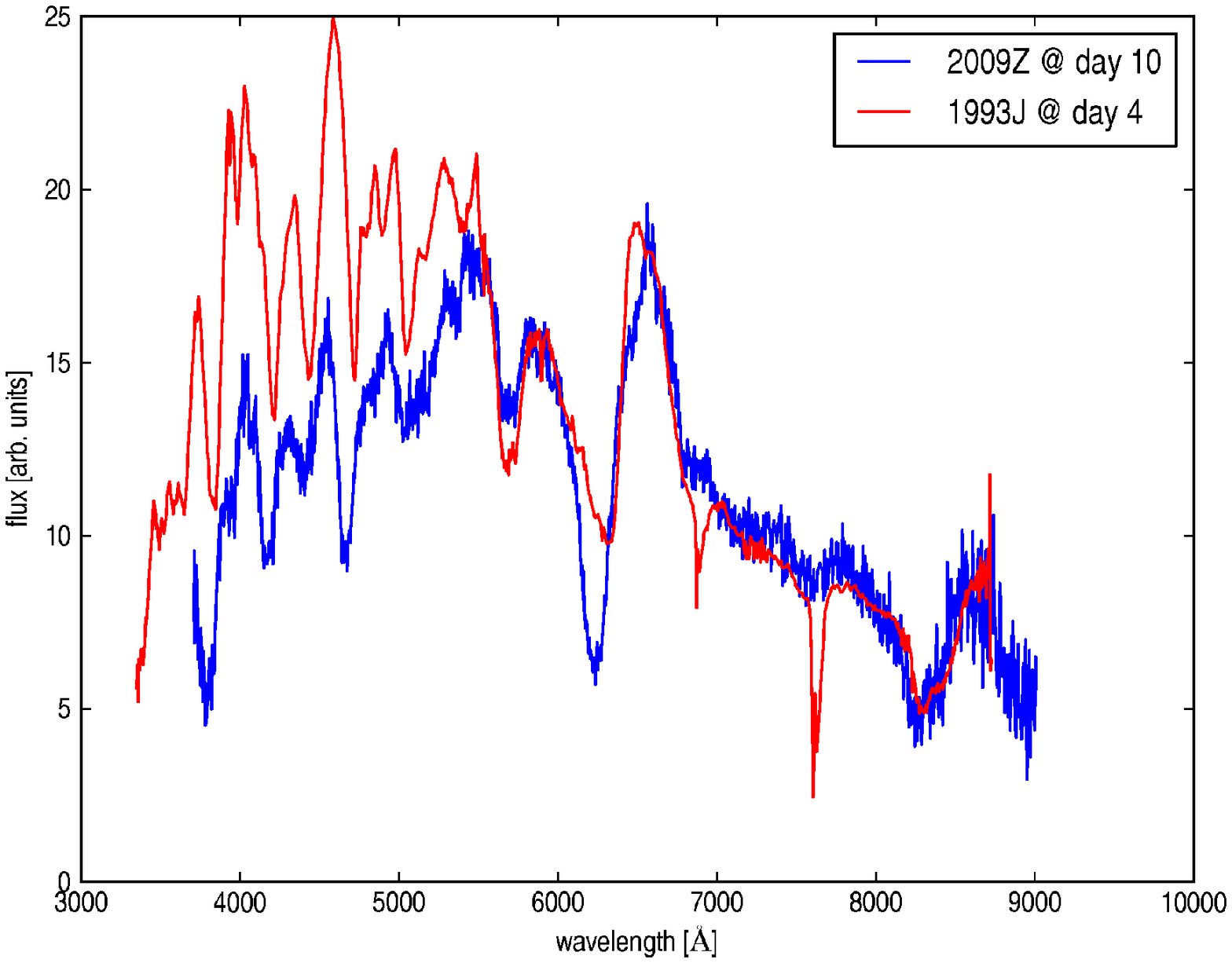}
  \includegraphics[width=0.49\textwidth]{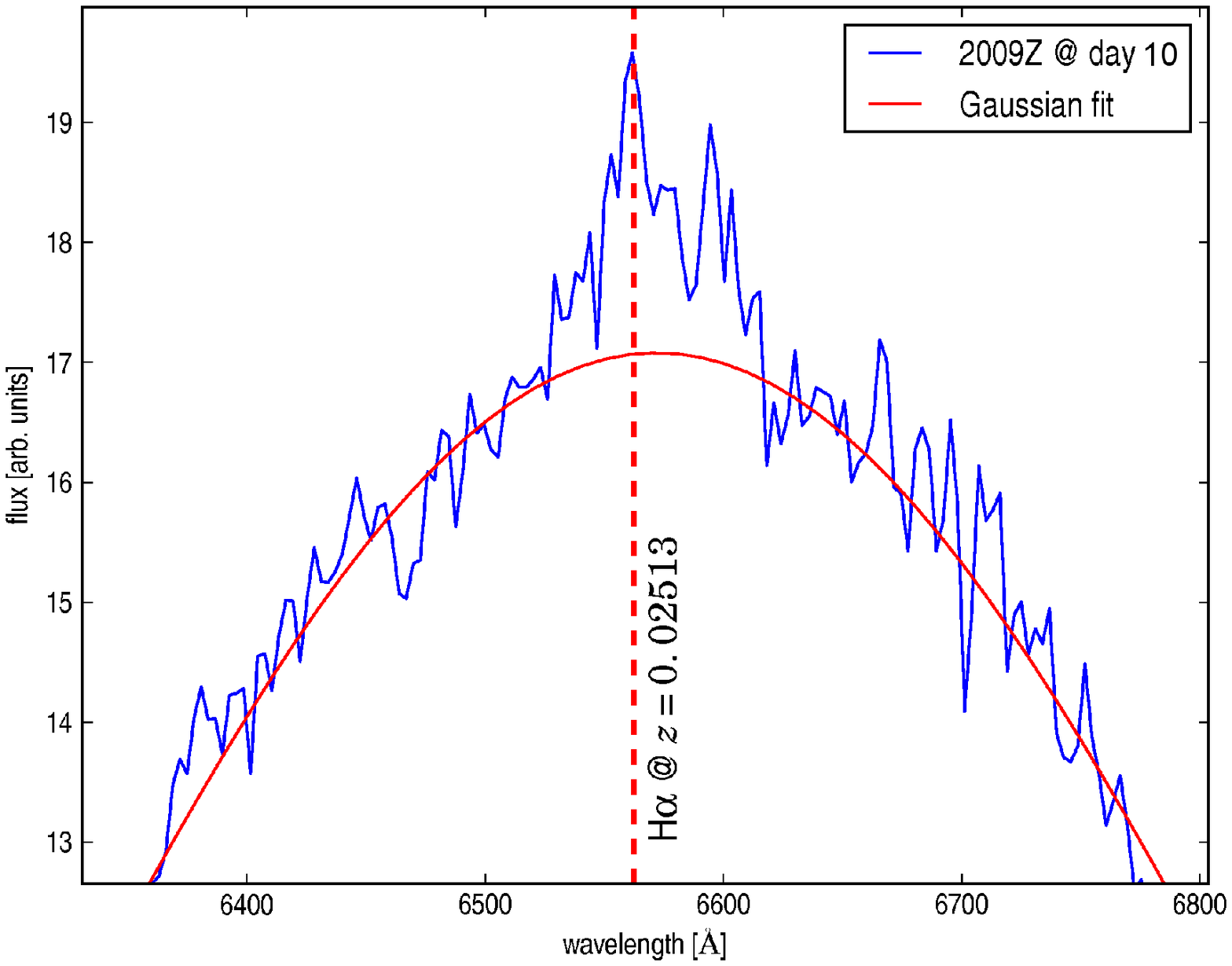}
  \caption{Left, an optical spectrum of SN\,2009Z (blue line) obtained 10 days after maximum light with the 2.5\,m du Pont Telescope at Las Campanas Observatory is shown. The spectrum is de-redshifted using $z=0.02513$ as derived from our H\,I observations (see Sect.~\ref{HI}). A similar spectrum of SN\,1993J, the archetype explosion for type IIb events, obtained 4 days after bolometric maximum \citep[see][]{Barbon1995}, is overlaid (red line) for comparison. Fluxes of both spectra were normalized at 6000\,$\AA$. In the right panel, the region around the H$\alpha$ line of SN~2009Z is shown. A Gaussian fit (red line) to this line reveals the presence of a small amount of H$\alpha$ emission ``on top'' of the broad supernova line. This additional H$\alpha$ flux is most likely to originate from the host galaxy N271.
    \vspace{3em}}
  \label{spectrum}
\end{figure*}

For the purpose of validating the redshift of the host galaxy of SN~2009Z, N271, we show an optical spectrum of SN~2009Z in Fig.~\ref{spectrum}. This spectrum was obtained 10 days after maximum light with the 2.5\,m du Pont Telescope at Las Campanas Observatory. We used $z=0.02513$ as derived from our H\,I observations of N271 presented in Sect.~\ref{HI} to de-redshift the spectrum and compare it to SN~1993J \citep{Barbon1995}, the archetype IIb event. Furthermore, we examined the prominent H$\alpha$ line to eventually detect H$\alpha$ emission from N271. Fitting a Gaussian to the broad H$\alpha$ line leaves a small ``cap'' on top of it (see right panel in Fig.~\ref{spectrum}). Since this ``cap'' exactly matches the redshift derived from the H\,I spectrum, we conclude that it is originating from N271 itself, adding to the H$\alpha$ emission of SN~2009Z.

\subsection{Host galaxy: archival data}
\begin{figure*}
  \centering
  \includegraphics[width=0.54\textwidth]{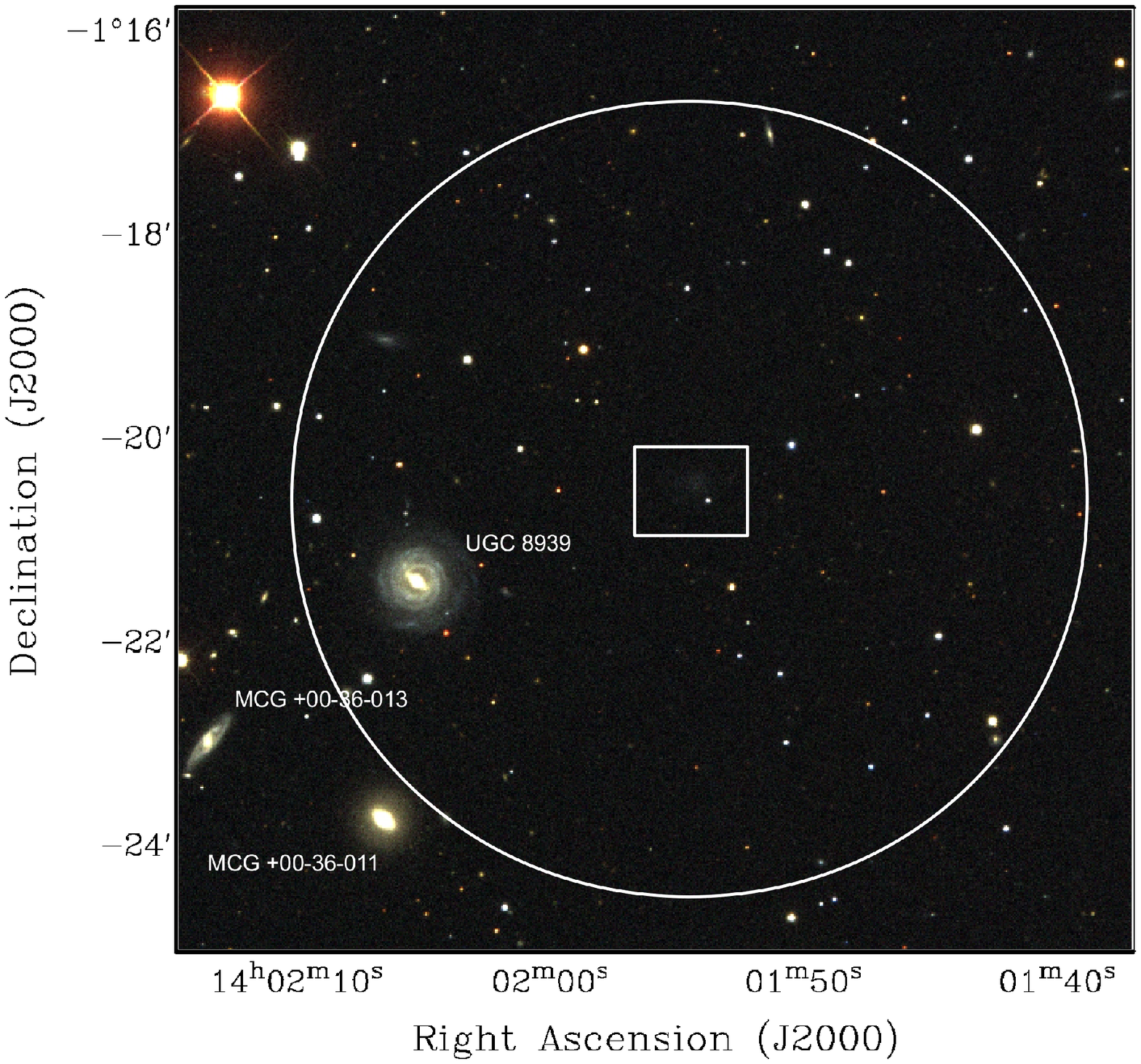}
  \includegraphics[width=0.44\textwidth]{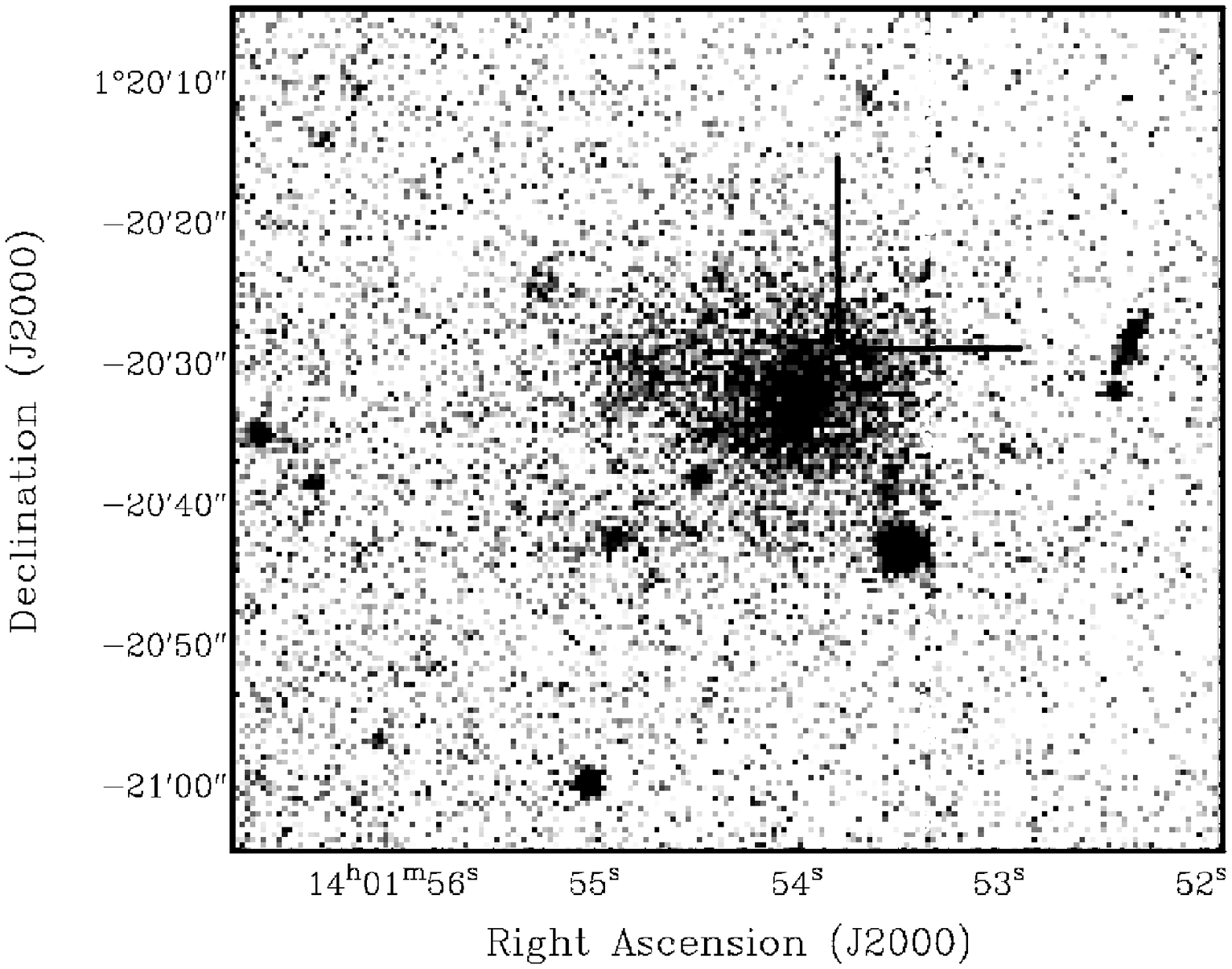}
  \caption{Left, color composite of SDSS $g$-, $r$-, and $i$-band images of N271 (located at the very center of the picture) and its surrounding. The circle indicates the Effelsberg half-power beam width at 21\,cm. Right, NTT $R$-band close-up of N271 (region of close-up highlighted in the left panel by a white box) with two ticks marking the position of SN~2009Z).
    \vspace{3em}}
  \label{SN2009Z}
\end{figure*}
\mk{We used archival photometric data on N271 from the SDSS and ESO archives. 
The left panel of Fig.~\ref{SN2009Z} shows a color SDSS image composed of  $g-$, $r-$ and $i$-band images of N271.
In addition, to enhance the accuracy  of the spectral energy distribution (SED) measurement of N271 (see Sec.~\ref{starform}), ultraviolet (UV) imaging data was obtained from 
the GALEX \citep{galex} database.
To ensure that both GALEX and SDSS magnitudes were comparable for the SED fitting process, 
 the GALEX flux densities in both the far ultraviolet (FUV at about 1500\,$\AA$) and near ultraviolet (NUV at about 2300\,$\AA$) bands were measured using the same aperture
as for the computation of the photometry from the SDSS images. 
The GALEX FUV image of N271 is shown in Fig.~\ref{Galex}. 
A journal of the complete photometric data set used in this work is given in Table~\ref{photo}.} Unfortunately, there is no infrared (IR) data for N271, neither in the 2MASS survey (due to a high flux limit) nor in the UKIDSS survey (which does not cover the location of N271 yet). 
Despite being detected in the 2dF survey \citep{Folkes1999},
 N271 has  no spectrum available with a  sufficient S/N to determine either its redshift or metallicity.

\begin{table}[t]
  \caption[]{Photometric data points used in this work and its sources.}
  \label{photo}
  $$
  \begin{tabular}{lccc}
    \hline
    \hline
    \noalign{\smallskip}
    Band & mag & mag error & Source \\
    \noalign{\smallskip}
    \hline
    \noalign{\smallskip}
    $FUV$ & 20.56 & 0.07 & GALEX AIS\tablefootmark{a} re-measured\\
    $NUV$ & 20.30 & 0.06 & GALEX AIS\tablefootmark{a} re-measured \\        
    $u$ & 20.39 & upper limit\tablefootmark{c} & SDSS DR6\tablefootmark{b}     \\
    $g$ & 18.71 & 0.03 & SDSS DR6\tablefootmark{b}     \\
    $r$ & 18.57 & 0.04 & SDSS DR6\tablefootmark{b}     \\
    $i$ & 18.09 & 0.04 & SDSS DR6\tablefootmark{b}     \\
   $z$ & 18.13 & upper limit\tablefootmark{c} & SDSS DR6\tablefootmark{b}     \\
    \noalign{\smallskip}
    \hline
  \end{tabular}\\
  $$
  \tablefootmark{a}{\cite{galex}, fluxes were re-extracted to ensure the same aperture radius as for the optical data.}\\
  \tablefootmark{b}{\cite{SDSS-DR6}}.\\
  \tablefootmark{c}{
Because the $u$- and $z$-band SDSS images are of low signal-to-noise (S/N), 
  no detection of  N271 could be made down to the 2.5 $\sigma$ level, hence
  these values are not used when fitting the SED. 
  }\\
\end{table}
As the SDSS images are neither sensitive enough nor provide sufficient spatial resolution to perform morphological analyses for such faint galaxies as N271, 
it was necessary to obtain additional deep imaging, particularly to determine the scale length of N271 and allow a precise measurement of its
surface brightness profile.
We therefore obtained from the ESO archive
  an $R$-band image taken on June 23, 2004 with EMMI mounted to the 3.6\,m New Technology Telescope (NTT). 
With an integration time of 300\,s this image is much deeper than those
 from the SDSS archive (52\,s with a 2.5\,m mirror).
The NTT image also benefits from excellent seeing conditions ($\sim$ 0$\farcs$6). The right panel of Fig.~\ref{SN2009Z} shows a close-up of the NTT image of N271 used for the scale length fitting described below.

 By assuming an exponential
surface brightness profile \citep[as is commonly done for LSB galaxies, see e.g.][]{oneil97}, we compute a central $B$-band surface brightness of $\mu_B\,=\,24.08\pm 0.13$\,mag\,arcsec$^{-2}$. For this calculation a scale length of $h_r\,=\,1.5$\,kpc was adopted as measured from the NTT image using the
  in IRAF\footnote{IRAF is distributed by the National Optical Astronomy Observatories, which are operated by the Association of Universities for Research in Astronomy, Inc., under cooperative agreement with the National Science Foundation.} task \texttt{ellipse}.
Adopting a magnitude cut definition for LSB galaxies either of 23 \,mag\,arcsec$^{-2}$ \citep{impeybothun} or 22\,mag\,arcsec$^{-2}$ \citep{mcgaugh95}, N271 is clearly a LSB galaxy.
\begin{figure}
  \centering
  \includegraphics[width=0.5\textwidth]{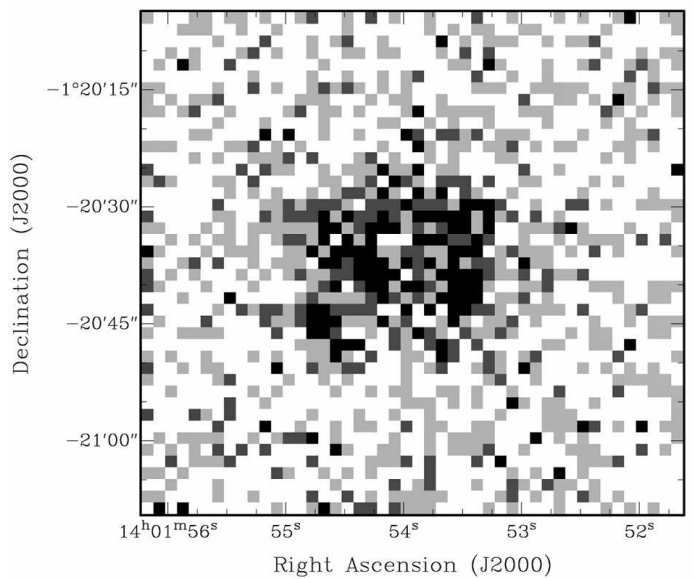}
  \caption{GALEX FUV image of N271. This image consists of approximately the same region as the NTT $R$-band image shown Fig.~\ref{SN2009Z}. 
  As one can clearly see, N271 looks very flocculent in the FUV, in particular when looking at the east and west edges of the galaxy whereas a central stripe has significantly less UV emission. This supports the idea described above that LSB galaxies have
   localized starforming regions in contrast to a starburst event across the entire galaxy.}
  \label{Galex}
\end{figure}

\subsection{H\,I spectroscopy}
\label{HI}
A H\,I spectrum of N271 was obtained with the 100\,m Effelsberg  Radio Telescope. 
This spectrum was used  to measure the redshift from the  21\,cm line, 
as well as to  determine the  H\,I mass content.
Therefore a 20\,MHz filter was used, spread over a
frequency region from\footnote{This range was
  chosen because it corresponds to the redshift of UGC 8939, with which we assumed N271 is associated.}
   1374\,MHz to 1394\,MHz, distributed over 16384
channels. This gave a spectral resolution of 1.22\,kHz or approximately
0.26\,km\,s$^{-1}$, well enough to separate even small velocity and
hence redshift differences. During the reduction  of the
spectrum, a binning (bin-width of  six channels) was performed to 
increase the S/N ratio.

The FWHM beam width at this wavelength is 8\,arcmin so we expect lines from more than one galaxy: in fact we find two clear H\,I emission lines (Fig.~\ref{effelsberg}). UGC~8939, whose redshift is already known \citep[$z$\,$=$\,0.0248,][]{fairall92}, we identify with the stronger, less redshifted line.
 One might expect a signal from MCG+00-36-011, but this galaxy has a redshift of 0.0249 and may blend with the signal from UGC~8939; also because it is located just outside the FWHM circle and because elliptical galaxies in general show much weaker H\,I emission than spirals or irregulars we conclude that this galaxy was entirely not detected or is at least only very slightly affecting the detection of UGC~8939. Therefore we identify the weaker of the two observed lines with N271.

We stress that there is only little risk of confusing the H\,I signals identified in Fig.\ref{effelsberg}. The shape of the emission line belonging to UGC~8939 matches exactly the expactations for an H\,I line of a face-on spiral, so a single line instead of a doulbe peaked profile which is typical only for edge-on spirals. The non-detections of the two galaxies MCG+00-36-011 and MCG+00-36-013 are very plausible, too, since the former is an elliptical which are known for having much smaller H\,I reservoirs than spirals and the latter one being pretty much outside the Effelsberg beam with an antenna response of only 0.2\% at this distance from the pointing center.

A baseline subtraction and Gaussian fits to the two lines were performed in order to measure the redshifts and H\,I masses, following the method of \cite{roberts62}. 
For N271, this yielded a peak radial velocity of 
v\,=\,7535$\pm$\,3\,km\,s$^{-1}$ relative to the rest frequency of 
neutral hydrogen\footnote{This velocity is calculated relative to the Local Standard of Rest (LSR). Please note that for the estimation of a cosmological redshift no peculiar velocity of N271 and no rotation correction for the MW were taken into account.}, corresponding to a redshift of $z$\,=\,0.02513$\pm$\,0.00001. This implies a 
luminosity distance $d_L\,=\,108.1\pm 0.4$\,Mpc. Note that this is a cosmologically determined distance based on the redshift of N271 and the $\Lambda$CDM cosmological model adopted in Sec.~\ref{Intro}. Therefore its error only reflects the measurement uncertainty of the redshift, errors due to a peculiar velocitiy that N271 may have or errors of the cosmological parameters were not taken into account.

From the stronger emission line of UGC~8939, we measure a H\,I mass of $8\,\,10^9\,M_{\odot}$, typical of a Sb spiral galaxy. For N271 we arrive at
\begin{equation}
M_{H\,I}\,=\,2.96\pm 0.12\,\,10^9\,M_{\odot},
\end{equation}
putting N271 within the (upper part of the) range of H\,I masses of dwarf galaxies \citep[see e.g.][]{Zwaan05}.
This, and its absolute $B$-band magnitude of $M_B\,=\,-16.22$, confirm that 
N271 is a LSB dwarf galaxy.

\begin{SCfigure*}
  \centering
  \includegraphics[width=0.8\textwidth]{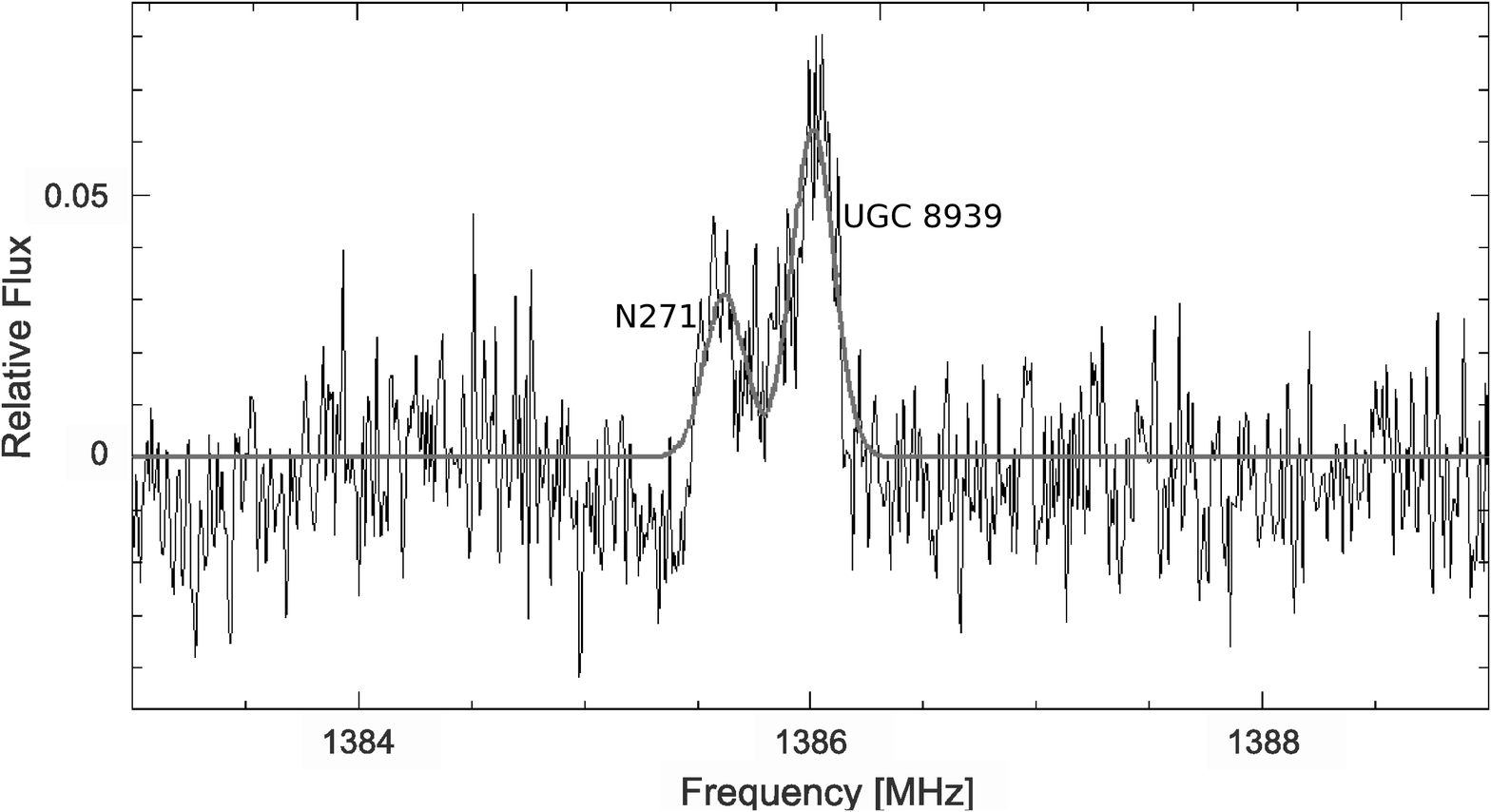}
  \caption{H\,I Effelsberg spectrum. The black
    line is the actual measurement, while the smooth grey line represents 
    Gaussian fits of the two detected emission lines after baseline subtraction. The galaxy identities are marked.
    Note that the flux is given in arbitrary units.}
  \label{effelsberg}
\end{SCfigure*}

\section{Inferred properties of N271}\label{properties}

\subsection{Stellar mass}
\label{stellarmass}
We now measure the stellar mass of N271 using the work of \cite{bell03} who connected the stellar mass
of a galaxy to measurable parameters assuming a \cite{Salpeter55} initial mass function (IMF). Based on the $g$- and $r$-band luminosities of N271, which are less affected by contemporary star formation than the $B$-band (classically used for this estimation), we
compute a stellar mass of
\begin{eqnarray}
  \log(M^*/L_r)\,&=&\,1.431(g-r)-0.022 \cr
  \rightarrow M^*\,&=&\,2.62\,\pm 1.77\,\,10^8\,M_{\odot}.
\end{eqnarray}
From this a gas mass fraction $f_g\,=\,M_{g}/(M_{g}+M^*)$ can be derived following
\cite{schombert01} of
\begin{eqnarray}
  f_g\,&=&\,\left(1+\frac{M^*}{\eta M_{H\,I}}\right)^{-1}\,=\,0.87 \pm 0.04,
\end{eqnarray}
where we have adopted $\eta\,=\,1.4$ as the inverse hydrogen mass fraction from \cite{vallenari05},
 which corresponds to solar
composition. 
Of the large survey of LSB dwarfs by \citet{schombert01}, only three galaxies have a comparably high gas fraction $f_g\,\approx\,0.9$, and the mean H\,I mass is around three times lower than that in N271. Also, only one percent of the sample have a central surface brightness as low as 24\,mag\,arcsec$^{-2}$. In summary, N271 is a rather extreme LSB dwarf specimen.

\subsection{Star formation history}\label{starform}
To characterize the  stellar population of N271 we used the available photometric data to fit SED templates of various galaxy types, using the publicly available SED fitting code \texttt{hyperz} by \cite{hyperz}. This package contains two sets of template SEDs, both of which we included in our fitting procedure. Firstly, the observationally generated templates by 
\cite{CWW}, CWW hereafter, include  E/S0, Sbc, Scd and Im galaxies. Secondly, the synthetically generated templates from the GISSEL library by 
Bruzual \& Charlot (1993; evol for short) include starburst, E, S0, Sa, Sb, Sc, Sd and Im galaxies, and allow an estimation of the age of the stellar population. 
The fitting process is a $\chi^2$ minimization technique which makes use of the measured redshift of $z_{fit}$\,=\,0.025.
  The two best-fit SED templates are shown in Fig. \ref{SEDs}.
\begin{figure*}
  \centering
  \includegraphics[width=0.48\textwidth]{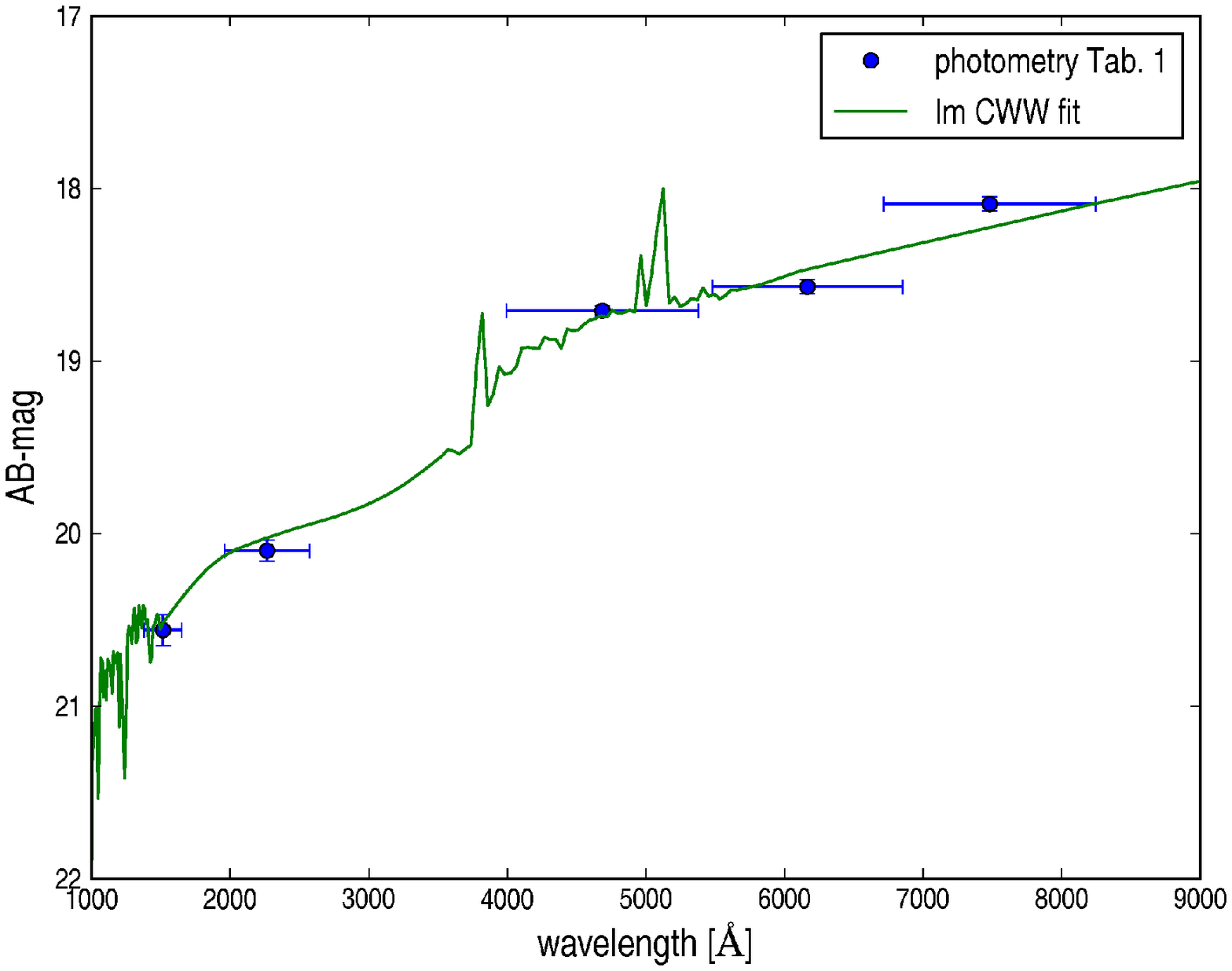}
  \includegraphics[width=0.48\textwidth]{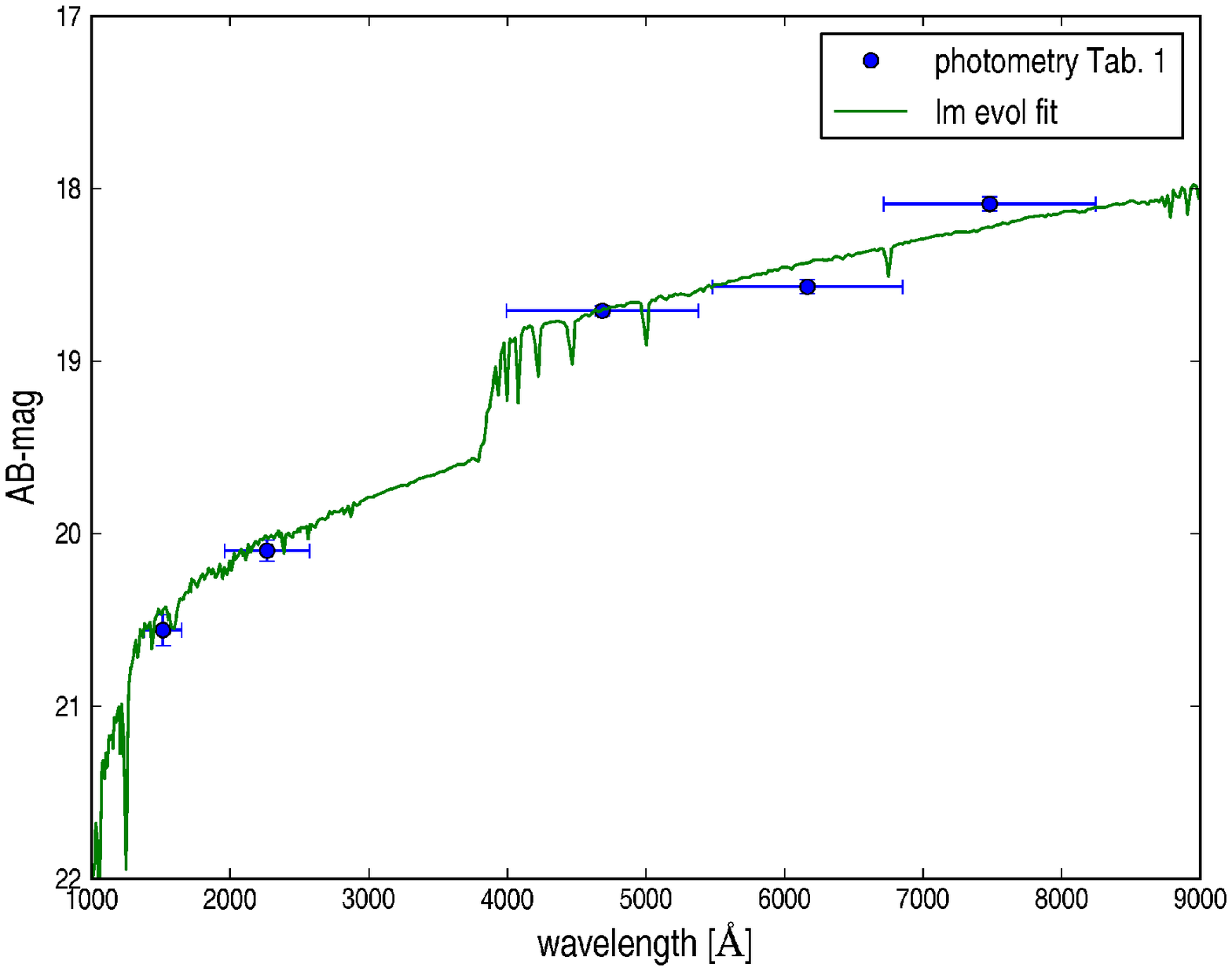}
  \caption{The two best-fit model SEDs computed with \texttt{hyperz}. {\it Left:} The best-fitting model ($\chi^2$\,=\,3.1) of an Im galaxy from the \cite{CWW} template set; {\it right:} The Im galaxy template ($\chi^2$\,=\,5.3) from the \cite{Bruzual93} synthetic template set which also allows an age estimation of the stellar population, which in this case is 360\,Myr. The corresponding extinction coefficient employed by the model is $A_V\,=\,1.0^m$. The x error bars correspond to the FWHM of the GALEX and SDSS bandpasses.}
  \label{SEDs}
\end{figure*}
The best-fit model favors an irregular galaxy from the empirically determined \cite{CWW} set. A second fit employing only the Im template from the synthetic \cite{Bruzual93} set was done, yielding an only slightly worse fit ($\chi^2_{evol}$\,=\,5.328 compared to $\chi^2_{CWW}$\,=\,3.128). From this, a stellar population age of $360\pm160$\,Myr was determined. 
This relatively young age \citep[see e.g.][for a characterization how young stellar populations influence their parent galaxies]{Li07} is in good agreement with the work by \cite{haberzettl08} who analyzed LSBs in the {\it Hubble} deep fields, resulting in the main finding that the stellar populations of LSBs tend to be younger than in comparable HSBs. This would either imply that N271 is currently undergoing its first major star formation event at all or at least has become active again after a longer phase of quiescence. This is also  emphasized by the high extinction $A_V\,=\,1.0^m$ employed by the best-fitting model. Deriving the current star formation rate of N271 using the FUV flux as measured using GALEX data ($L_{FUV}\,=\,3.00\pm 0.05\,\,10^{26}\,$erg\,s$^{-1}$\,Hz$^{-1}$) as a star formation tracer, following the calibration of \cite{Kennicutt98} and \cite{Madau1996}, one gets a star formation rate of:
\begin{eqnarray}
  SFR_{FUV}\,&=&\,0.0420\pm 0.033\,M_{\odot}\,\rm{yr}^{-1}, \cr
  SFR_{FUV,corrected}\,&=&\,0.44\pm 0.34\,M_{\odot}\,\rm{yr}^{-1}.
\end{eqnarray}
The corrected SFR accounts for extinction in the galaxy itself which is most important since UV wavelengths are extremely affected by dust attenuation. To correct the FUV flux for this extinction, we adopted $A_V\,=\,1.0^m$ from the SED-fit, yielding an attenuation at 1516\,$\AA$  (effective wavelength of the GALEX FUV band) of $A_{FUV}\,=\,2.5^m$ using the \cite{Calzetti00} extinction law. Note that $FUV$-derived SFRs are known to be notoriously affected by the amount of internal extinction and the reddening curve adopted for the computation of $A_{FUV}$ and that therefore the value actually derived as SFR has to be treated with care (see. Sec.~\ref{comparison} for a closer discussion). 

Nevertheless, we come to the conclusion that a star formation rate of a few 0.1\,$M_{\odot}\,\rm{yr^{-1}}$ supports the idea that N271 is currently undergoing a major star formation episode, considering that it is a LSB dwarf galaxy. For comparison we note that normal HSB spiral galaxies such as the Milky Way typically show star formation rates of $\sim\,1\,M_{\odot}\,\rm{yr}^{-1}$, only a factor of a few higher than that of N271, although it is about 100 times more massive.

\subsection{Metallicity}
\label{metallicity}
{\mk Given this relatively high SFR, one might wonder whether the currently ongoing star-formation event is the first one in the history of N271 or whether there has been previous star-formation activity. By estimating the metallicity of N271, we shall see that the latter of these two possibilities is much more likely. 
Due to the lack  of a sufficiently high S/N spectrum, we estimated the metallicity using  the rough metallicity -- luminosity relationship, as calibrated by \cite{Pilyugin01} for dwarf irregular galaxies.} Based on N271's absolute $B$-band magnitude calculated using the \cite{fukugita96} conversation equations between SDSS and Johnson/Kron-Cousins bandpasses, $M_B\,=\,-16.22$, we obtain an oxygen-related gas phase metallicity of
\begin{eqnarray}\label{uncorrected}
  12+\log\left(\frac{\rm{O}}{\rm{H}}\right)\,&=&\,8.05\pm 0.67, \cr
  12+\log\left(\frac{\rm{O}}{\rm{H}}\right)_{corrected}\,&=&\,8.24\pm 0.70. 
\end{eqnarray}
As before, we used an extinction of $A_V\,=\,1.0^m$ as suggested by the SED-fit and a \cite{Calzetti00} extinction law to compute an extinction-corrected $B$-band absolute magnitude of $M_{B,corrected}\,=\,-17.67$. Despite the relatively large error  
associated with the metallicity\footnote{This large error arises due to the large intrinsic scatter of the well known relationship between metallicity and luminosity, see for example \cite{Pilyugin01}, their Fig.~2.} inferred from the coarse relation between metallicity and luminosity, one has also to bear in mind that N271 was demonstrated to be an extreme example of a dwarf galaxy, hence may fall off the calibration by \citeauthor{Pilyugin01} at an even larger fraction. But since the galaxy is quite faint so that there are no optical spectra available from which a more accurate metallicity could be derived, we decided to adopt this value with the corresponding errors. With that, one cannot regard N271 to be a metal-poor galaxy, particularly if one compares it to other low-luminosity dwarf irregulars such as the Large and Small Magellanic Clouds ($12+\log\left(\rm{O}/\rm{H}\right)\,=\,8.50$ and $8.09$ respectively which also fall very well on the metallicity -- luminosity relation). This implies, even when only considering the uncorrected value in (\ref{uncorrected}), that earlier SNe must have occurred in N271 to enrich the ISM with metals.
Assuming a constant SFR throughout the entire 360\,Myr of age of the fitted stellar population of N271 we come up with a total mass of stars formed of $M^*_{tot}\,=\,1.6\,\,10^8$ $M_{\odot}$. This is  in good agreement with the mass-to-light ratio determined stellar mass as well as with the stellar mass derived from a Bayesian approach as outlined in Sect.~\ref{comparison}. All together, this
underlines the picture that star formation in LSB galaxies occurs in small distinct bursts 
that are well separated in time because the bulk of the
 stellar content of N271 seems to have been produced during the current burst.

\subsection{Testing the derived parameters} 
\label{comparison}
To test the reliability of the galaxy's parameters derived so far using a variety of well-known scaling
relations, we also performed a more sophisticated SED fit which follows a Bayesian approach. We make use of a large library of model SEDs obtained by convolving \cite{Bruzual03} simple stellar populations of different metallicities with Monte Carlo star formation histories and dust attenuations. For dust attenuation we adopt in this case the \cite{Charlot2000} two-component model, regulated by the total effective optical depth $\tau_V$ affecting stars younger than $10^7$\,yr and the fraction $\mu$ contributed by the ISM. As a result the dust attenuation curve is not constant in time and is not a simple power law for composite stellar populations (as opposed to the \cite{Calzetti00} attenuation law adopted above).

To derive galaxy's physical parameters such as stellar mass, dust attenuation, mean light-weighted age and SFR, we compare the galaxy SED to all the SEDs in the model library and build the probability density function of each parameter. The advantage of this approach is that it provides a robust estimate of the uncertainties in the derived parameters coming both from observational uncertainties and model degeneracies. It is however more sensitive to the adopted prior distribution of the model parameters.

\begin{table}[t]
  \caption[]{Synopsis of SED-fitting derived parameters.}
  \label{SED}
  \begin{tabular}{lccc}
    \hline
    \hline
    \noalign{\smallskip}
     & classic & classic & Bayesian \\
     & Salpeter IMF & Chabrier IMF & Chabrier IMF \\
    \noalign{\smallskip}
    \hline
    \noalign{\smallskip}
    $A_V$ [mag] & $1.00\pm0.15$ & & $0.93\pm0.51$ \\
    $A_{FUV}$ [mag] & $2.52\pm0.2$ & & $1.26\pm0.55$ \\
    $M^*$ [$10^8\,M_{\odot}$] & $2.62\pm1.77$ & $1.54\pm1.04$  & $1.98\pm1.39$ \\ 
    $SFR$ [$M_{\odot}\,\rm{yr^{-1}}$] & $0.44\pm0.34$ & $0.29\pm0.23$ & $0.09\pm0.04$ \\
    age [Myr] & $360\pm160$ & & $960\pm1359$ \\
    \hline
  \end{tabular}
\end{table}
 
The parameters derived both with the ``classical'' method and with the Bayesian approach are summarized in Tab.~\ref{SED}. Note that the ``classical'' parameters were calculated adopting a \cite{Salpeter55} Initial Mass Function (IMF) whereas the Bayesian calculation relies on a \cite{Chabrier2003} IMF. While the choice between the two IMFs does not affect the color evolution, hence color-derived stellar population properties such as age and dust attenuation, it affects integrated quantities such as stellar mass and SFR. Based on \cite{Bruzual03} models, we estimate that $M_{Salp}\approx1.7\,M_{Chab}$ and
$SFR_{Salp}\approx1.5\,SFR_{Chab}$ and adopt these conversions for comparison between the two approaches in Tab~\ref{SED}. Both the stellar mass and the attenuation in the optical derived with the two methods agree very well within the combined uncertainties. The dust attenuation in the FUV is instead quite different, most likely as a result of the different attenuation laws adopted in the two cases. This affects mostly the SFR estimate, which decreases by a factor of $\sim3$ if we adopt $A_{FUV}\,=\,1.26^m$ instead of $A_{FUV}\,=\,2.5^m$ to correct the FUV luminosity. We note though that by  adopting $A_{FUV}\,=\,1.26^m$ the SFR estimated directly from the UV luminosity using the \cite{Kennicutt98} formula agrees very well with the one derived by the Bayesian SED fitting. Because of the age--dust degeneracy the difference in dust attenuation is also somewhat reflected in the estimated stellar age, which is higher in the Bayesian approach.

As a whole these results point to a low-mass galaxy with a young stellar population having ongoing star formation at a level of at least $0.1\,M_{\odot}\,\rm{yr^{-1}}$ and a fair amount of dust attenuation.

\section{Discussion}\label{discussion}
The progenitors of SNe~IIb are believed to be massive single stars that  
have lost much of their hydrogen envelope \citep{Woosley1993} or  
 massive evolved stars in a binary system \citep[e.g.][]{Crockett2008,Thielemann1996, Woosley1995,Shigeyama1990}. In particular, the archetype of such SNe, SN 1993J was identified to have had a massive binary companion of $14\,M_{\odot}$ \citep{Maund2004}.
Type IIb SNe therefore 
demonstrate the presence of massive stars, so the case of SN~2009Z contradicts the long-held belief that LSB galaxies contain only low-mass stars, but corresponds to the findings of  
\cite{mattsson07}, that the initial mass functions (IMFs)
of LSB galaxies do extend to high mass.

The extremely high gas mass fraction is a strong hint that this type of galaxy is inefficient in star formation \citep[e.g.][]{schombert90}.
 Our star-formation rate estimate in N271 of 0.44 $\,M_{\odot}\,\rm{yr}^{-1}$ is larger than typical LSB galaxies, which lie in the range 0.02 to 0.2 $\,M_{\odot}\,\rm{yr}^{-1}$ \citep{vandenhoek00}, and almost comparable to that of normal HSB spirals.
  In addition its metallicity of $12+\log\left(\frac{\rm{O}}{\rm{H}}\right)_{corrected}\,=\,8.24\pm0.70$ is higher than typical LSB dwarfs, but is normal for HSB galaxies of comparable mass \citep{Pilyugin07}. 
  
 A star formation history of LSB galaxies that includes the existence of short (a few 100\,Myr) bursts separated by longer quiescent periods is preferred by many authors \citep{schombert01,boissier03,vallenari05,boissier08}; these starbursts are too short-lived to transform the galaxy to HSB.  In N271 this scenario looks very likely; the progenitor of this core-collapse SN presumably formed in the most recent starburst. 
   Furthermore, it may be related to the finding by Grunden et al. (in prep.) that the ratio of core-collapse to thermonuclear SNe is two times higher in LSB galaxies than in HSB galaxies.

SNe in dwarf galaxies have recently become a heavily discussed topic. It is informative to compare this SN in a LSB dwarf to SNe in other, both LSB and non-LSB, dwarfs. Using the first compilation of 72 SNe from the Palomar Transient Factory (PTF), \cite{Arcavi2010} analyzed statistics of CCSNe in dwarfs and giant galaxies. They found a significant excess of Type IIb events in dwarfs (defined as $M_r$\,$>$\,$-$18; N271 has $M_r$\,=\,$-$16.6), which they mostly consider to be a consequence of the lower metallicity in their dwarf sample: metal rich stars have strong winds and hence mass loss, so they explode as Type~Ic SNe, whereas metal-poor stars produce Type~Ib or Type~IIb events.
Given the coarse metallicity estimate of N271, SN~2009Z stands out considering this hypothesis because its host galaxy exhibits a similar metallicity as hosts of typical SNe~Ic associated with a GRB. In contrast, for a host of a IIb event, its metallicity is fairly high. This could be due to the LSB nature of N271 since \cite{Lee2004}, who analyzed LSB galaxies in terms of their IMF, found that LSBs could be fitted best by a \cite{Salpeter55} IMF with a significantly steeper exponent at the low-mass end (about twice the standard value), so they argue for a bottom-heavy IMF in LSBs which would lead to the formation of stars mostly well beyond the 8$\,M_{\odot}$ limit for CCSNe \citep[see][]{smartt09,Heger2003}. The case of SN~2009Z then clearly demonstrates that at least intermediate-mass star formation (as IIb event, the progenitor of SN~2009Z must have had at least 30$\,M_{\odot}$ as single star or 15$\,M_{\odot}$ when member of a binary system) does happen in LSBs, too. However, we want to point out once more that because of the large error bars of our inferred metallicity (see Sect. \ref{metallicity}, this conclusion has to be treated with care.

\begin{figure}
  \centering
  \includegraphics[width=0.5\textwidth]{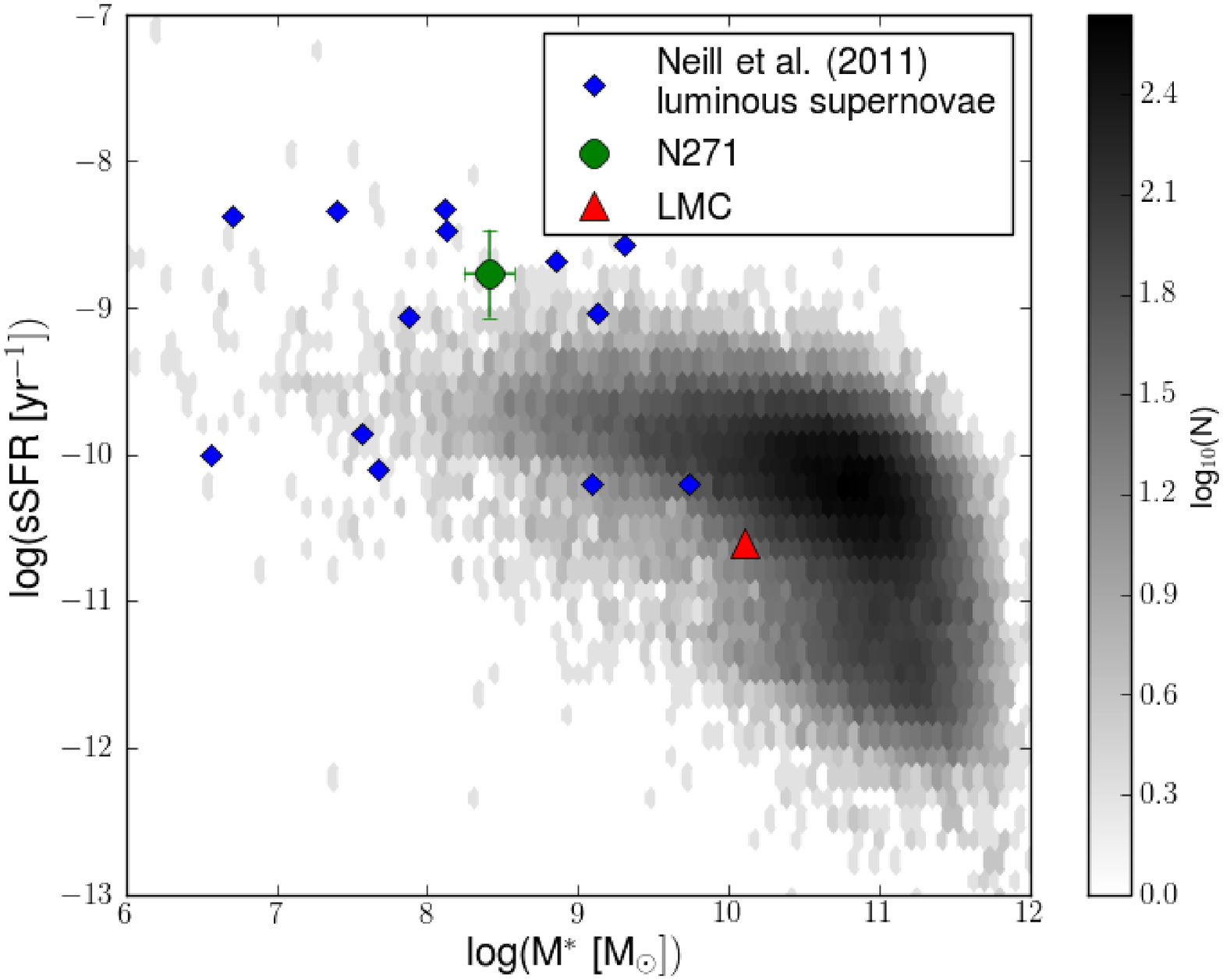}
  \caption{Specific star formation rate (sSFR) vs. stellar mass. The sample of hosts of luminous (M$_V\,<$\,-21) supernovae of \citet{Neill2011} (see their Fig.~3) is plotted together with N271 and the LMC (values from \citealt{Westerlund1997} and \citealt{Harris2009}) and  a sample of about 60,000 SDSS-GALEX galaxies from \cite{Wyder2007} relying on stellar masses and star formation rates deduced by \cite{Kauffmann2003} and \cite{Brinchmann2004}. Clearly, N271 fits quite well to the ``extreme'' hosts although SN~2009Z was not a very luminous event.}
  \label{LSNe}
\end{figure}

Although SN~2009Z was not a luminous event according to the definition of \cite{Neill2011}, i.e. peak M$_V\,<$\,$-$21, N271 falls well within the definition of having an ``extreme'' host galaxy that  \citeauthor{Neill2011} employ for the hosts of 13 luminous SNe.
 For this classification, they looked for the specific star formation rate (sSFR, defined as sSFR\,=\,SFR/M$^*$) of their luminous-SN host galaxies and found that most of them were very blue dwarfs with low stellar masses and high sSFRs. The sSFR of N271 is 1.71$\,\,10^{-9}$yr$^{-1}$, 
 which as we can see from Fig.~\ref{LSNe}, is well within the range of the hosts of these 
 luminous SNe. In accordance with the authors cited above, \citeauthor{Neill2011} 
 invoke metallicity to explain the correlation between faint, blue dwarf hosts and luminous SN events. Specifically, they argue that at higher metallicity, massive stars suffer much greater wind mass loss and that only in metal-poor galaxies one should expect to find the very massive ($>$\,100\,$M_{\odot}$) progenitors required to produce luminuous SNe \citep{Neill2011}.

Surveys of large areas of sky (of the order of a few thousand square degrees) with search cadences of a few days are discovering large numbers of SNe in low-metallicity galaxies. The Palomar Transient Factory now has a large sample of such SNe at low redshift and the relative rates are surprising \citep{Arcavi2010}. 
The Pan-STARRS1 survey is searching for low-z SNe in the 3Pi faint galaxy survey \citep{Valenti2010,Young2010}, too, but has also has found high-z ultraluminous
SNe at z=0.9 in dwarf galaxies \citep{Chomiuk2011}.

The largest sample of supernovae investigated in the context of their host galaxies was presented by \cite{Prieto2008}. They matched the SAI catalog to the SDSS DR4 value-added catalog by \cite{Kauffmann2003} to get metallicity information for the hosts of about 120 supernova events of all types. Their main finding that SN Ib/c seem to be more abundant in metal-rich galaxies while SN II seem to occur more often in metal-poor ones, also supports the argumentation outlined in this paper. They also matchd their supernova sample to pure SDSS image data to go to fainter host galaxies. This resulted in the finding that luminous supernovae tend to appear in faint hosts, as for instance the hypernova-like event SN~2007bg which happened in an extreme dwarf of $M_B=-12.4\pm0.6$\footnote{Despite this very low luminosity in the $B$ band, \cite{Young2010} determined the metallicity of the host to be $12+\log\left(\frac{\rm{O}}{\rm{H}}\right)\,=\,8.18\pm 0.17$, more than expected for such an extreme dwarf galaxy.}, one of the faintest SN hosts ever observed \citep{Young2010}.

The findings concerning the star formation history of N271 and its current stellar content
 could be related by taking into account the work by \cite{Rosenbaum2004} who analyzed the environment of LSB galaxies in the SDSS early data release. They find that LSBs are, unlike HSBs, often found in less dense environments or even in void structures. Therefore they undergo fewer interactions with other galaxies, which are known to trigger star formation.

\section{Conclusions}\label{conc}
We have investigated the dwarf galaxy N271 which is the host of the Type IIb SN~2009Z. It is a low surface brightness (LSB) galaxy with central surface brightness $\mu_B\,=\,24.08\pm 0.13$\,mag\,arcsec$^{-2}$. 
 Using a 21cm spectrum obtained with the Effelsberg Radio Telescope we measured a redshift of $z$\,=\,0.0251 and an H\,I mass of $\,2.96 \pm 0.12\,\,10^9\,M_{\odot}$.
Using SDSS $g$- and $r$-band magnitudes to estimate a mass-to-light ratio and therefore a stellar mass, we arrive at the rather high gas mass fraction of $f_g\,=\,0.87 \pm 0.04$.

SED-fitting using UV (GALEX) and optical (SDSS) data points yields a best-fit model of an irregular galaxy with a relatively young stellar population of age 360\,Myr. This is in good agreement with the (extinction corrected) FUV-derived star formation rate of $0.44\,M_{\odot}$\,yr$^{-1}$, which is somewhat higher than typical LSB values.
 This picture of N271 currently witnessing a starburst event  is supported by its relatively high metallicity of $12+\log\left(\frac{\rm{O}}{\rm{H}}\right)_{corrected}\,=\,8.24\pm 0.70$, comparable to the Magellanic clouds, implying metal-enrichment from previous bursts. Such distinct bursts may be  a common phenomenon in  LSB dwarf galaxies.

We conclude that LSB galaxies do not represent a completely alternative evolutionary path from HSB galaxies but rather are in a certain evolutionary state, as proposed e. g. by \cite{haberzettl08} and \cite{vandenhoek00}. The fact that there is high-mass star formation in LSBs (as shown by SN~2009Z) clearly demonstrates that the old picture of LSB galaxies being trapped in a low metallicity/low star-formation ``cage'' must be revised. More likely is that these galaxies are going through an LSB phase but at some time evolve into normal HSB galaxies.
 Whether this LSB phase is long or short may depend on the strength of the small bursts of star formation, more precisely if one of those bursts is strong enough to permanently transform the LSB into an HSB galaxy. Further exploration is needed to investigate this phase for HSB galaxies: Whether present-day HSBs ever went through a significant LSB phase is not yet clear.
  Finally, further work is needed regarding the apparent correlation between SN type and host galaxy type.

\begin{acknowledgements}
  We like to thank our referee, S.~J. Smartt, for all the helpful comments, especially on contemporary literature from the ``supernova side'', and further suggestions he made which substantially improved this paper.\\
  DARK is funded by the Danish NSF. We thank Norbert Langer for financial support.
  Funding for the SDSS and SDSS-II has been provided by the Alfred P. Sloan
  Foundation, the Participating Institutions, the National Science Foundation,
  the U.S. Department of Energy, the National Aeronautics and Space
  Administration, the Japanese Monbukagakusho, the Max Planck Society, and the
  Higher Education Funding Council for England. The SDSS Web Site is
  http://www.sdss.org/.\\
  Based on observations made with ESO Telescopes at the La Silla or Paranal Observatories under programme ID 073.A-0503(A).
\end{acknowledgements}

\bibliography{SN2009Z}
\bibliographystyle{aa}

\end{document}